\documentclass[%
 reprint,
superscriptaddress,
showpacs,
showkeys,
 amsmath,amssymb,
prl,
]{revtex4-1}

\usepackage{graphicx}
\usepackage{dcolumn}
\usepackage{bm}
\usepackage{esvect}
\usepackage{accents}

\newcommand{\be}{\begin{equation}}
\newcommand{\ee}{\end{equation}}
\newcommand{\ba}{\begin{eqnarray}}
\newcommand{\ea}{\end{eqnarray}}

\newcommand{\jcap}{JCAP}

\begin{document}

\preprint{APS/123-QED}

\title{MICROSCOPE mission: first constraints on the violation of the weak equivalence principle by a light scalar dilaton}

\author{Joel Berg\'e}
 \email{joel.berge@onera.fr}
\affiliation{DPHY, ONERA, Universit\'e Paris Saclay, F-92322 Ch\^atillon, France}

\author{Philippe Brax}
\affiliation{Institut de Physique Th\'eorique, Universit\'e Paris-Saclay, CEA, CNRS, F-91191 Gif-sur-Yvette Cedex, France}

 \author{Gilles M\'etris}
  \affiliation{Universit\'e C\^ote d'Azur, Observatoire de la C\^ote d'Azur, CNRS, IRD, G\'eoazur, 250 avenue Albert Einstein, F-06560 Valbonne, France}

\author{Martin Pernot-Borr\`as}
\affiliation{DPHY, ONERA, Universit\'e Paris Saclay, F-92322 Ch\^atillon, France}
\affiliation{Sorbonne Universit\'e, CNRS, Institut d'Astrophysique de Paris, IAP, F-75014 Paris, France}

\author{Pierre Touboul}
\affiliation{DPHY, ONERA, Universit\'e Paris Saclay, F-92322 Ch\^atillon, France}

\author{Jean-Philippe Uzan}
\affiliation{Institut d'Astrophysique de Paris, CNRS UMR 7095,
Universit\'e Pierre \& Marie Curie - Paris VI, 98 bis Bd Arago, 75014 Paris, France}
\affiliation{Sorbonne Universit\'es, Institut Lagrange de Paris, 98 bis, Bd Arago, 75014 Paris, France}

\date{\today}

\begin{abstract}
The existence of a light or massive scalar field with a coupling to matter weaker than gravitational strength is a possible source of violation of the weak equivalence principle. 
We use the first results on the E\"otv\"os parameter by the MICROSCOPE  experiment to set new constraints on such scalar fields. For a massive scalar field of mass smaller than $10^{-12}$~eV (i.e. range larger than a few $10^5$~m) we improve existing constraints by one order of magnitude to $|\alpha|<10^{-11}$ if the scalar field couples to the baryon number and to $|\alpha|<10^{-12}$ if the scalar field couples to the difference between the baryon and the lepton numbers. We also consider a model describing the coupling of a generic dilaton to the standard matter fields with five parameters, for a light field: we find that for masses smaller than $10^{-12}$eV, the constraints on the dilaton coupling parameters are improved by one order of magnitude compared to previous equivalence principle tests.
\end{abstract}

\pacs{04.50.Kd, 07.87.+v, 04.80.Cc}
\keywords{Experimental test of gravitational theories}
\maketitle



Scalar-tensor theories are a wide class of gravity theories that contain general relativity~\cite{will14}. In the Newtonian limit, they imply the existence of a fifth force, that can be well-described  by a Yukawa deviation to Newtonian gravity. Its range depends mostly on the mass of the scalar field and  can vary from sub-millimetric to cosmological scales~\cite{fischbach86, fischbach99}. It has so far been constrained on all scales from a few microns to the largest scales of the Universe (see e.g. Refs.\cite{will14, adelberger03, jain10}).

This new force may or may not be composition-dependent. A non-universal coupling implies both a violation of the weak equivalence principle (WEP) and a variation of the fundamental constants \cite{jpu1,jpu2}. The former effect has already been exploited by the E\"ot-Wash group to bring the current best constraints on Yukawa-type interactions and on light dilaton interactions \cite{schlamminger08, wagner12, damour12}, while the latter allows one to set constraints on cosmological to local scales~\cite{jpu3}.

The MICROSCOPE satellite aims to constrain the WEP in space \cite{touboul12, berge15a} by measuring the E\"otv\"os parameter, defined as the normalized difference of acceleration between two bodies $i$ and $j$ in the same gravity field, $\eta=\left(\Delta a/a\right)_{ij}=2\vert \vv a_i -\vv a_j\vert/\vert \vv a_i +\vv a_j\vert$. First results~\cite{touboul17a} give 
\begin{equation}\label{e.etaMicro}
\eta = \left(-1 \pm 27\right) \times 10^{-15}
\end{equation}
at a 2-$\sigma$ confidence level.
MICROSCOPE tests the WEP by finely monitoring the difference of acceleration of freely-falling test masses of different composition (Platinum and Titanium) as they orbit the Earth, measured along the principal axis of the (cylindrical) test masses. The measurement equation is given e.g. in \cite{touboul17a} as $a_{\rm Pt} - a_{\rm Ti} = g_x \eta + f(\vv{p}, n)$, where $g_x$ is the projection of the Earth gravity field onto the axis of the test and $f(\vv{p}, n)$ is a function of the instrumental and environmental parameters and measurement noise.

The constraint~(\ref{e.etaMicro}) was obtained after analyzing only one measurement session; therefore, the error bars should be considered as the largest that can be expected from the whole MICROSCOPE mission. The statistical error is expected to decrease with increasing data and with the refinement of the data analysis by the end of the mission in 2018. In the meantime, this new constraint of the WEP can already be used to set new bounds on fifth force characteristics.
This letter focuses on  the implications of the first results of MICROSCOPE  for an interaction between matter and a light dilaton.\\

\noindent{\it Scalar fifth force}. The existence of a light scalar field $\phi$ modifies the Newtonian interaction between two bodies $i$ and $j$ of masses $m_i$ and $m_j$ by a Yukawa coupling~\cite{adelberger03, DGef} \footnote{see Supplemental Material for a derivation of the equation}
\begin{equation} \label{eq_yukawa}
V_{ij}(r) = -\frac{Gm_im_j}{r} \left( 1 + \alpha_{ij} \hbox{e}^{-r/\lambda}\right).
\end{equation}
The scalar coupling to matter $\alpha_{ij}$ can be decomposed as the product $\alpha_i\alpha_j$ of the scalar couplings to matter for each test-body measured by the dimensionless factors (e.g. \cite{schimd05})
\begin{equation}\label{e.alphai}
 \alpha_i\equiv\frac{\partial\ln m_i/M_P}{\partial \phi/M_P}
\end{equation}
with $M_P^{-1}=\sqrt{4\pi G}$ the Planck mass. The range $\lambda$ of the Yukawa interaction is related to the mass of the field by $\lambda=\hbar/m_\phi c$. The amplitude of the WEP violation is related to the presence of a scalar field that does not couple universally to all forms of energy, contrary to general relativity. The magnitude  of the scalar force varies from element to element and is characterized by $\alpha_i(\phi)$ which requires the determination of $m_i(\phi)$ and thus the specification of the couplings of the scalar field to the standard model fields. Any dynamics or gradient of this scalar fields thus induce a spatial dependence of the fundamental constants~\cite{jpu1,jpu2}.  For two test masses in the external field of a body $E$, the E\"otv\"os parameter reduces to
\begin{equation}\label{e.eta1}
\eta = \frac{(\alpha_i-\alpha_j)\alpha_E}{1+\frac{1}{2}(\alpha_i+\alpha_j)\alpha_E}
      \simeq (\alpha_i-\alpha_j)\alpha_E.
\end{equation}
In order to set constraints, we need to specify the couplings of the field to matter as well as its masses.\\

\begin{figure}
\includegraphics[width=0.45\textwidth]{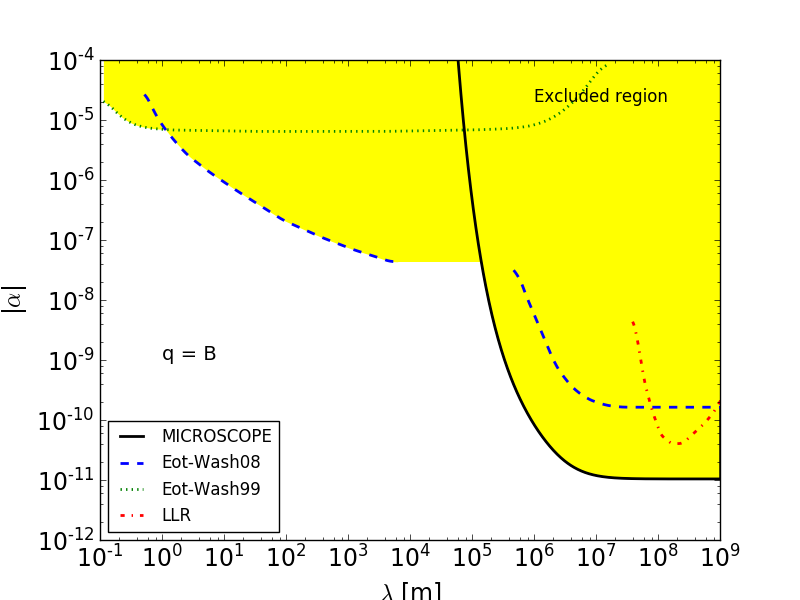}
\caption{Constraints on the Yukawa potential parameters ($\alpha, \lambda$) with $q=B$. The excluded region is shown in yellow and compared to  earlier constraints from Ref.~\cite{smith00} (dotted), Ref.~\cite{schlamminger08} (dashed) and Refs.~\cite{williams04, talmadge88} (dot-dashed). MICROSCOPE (solid line) improves on the E\"ot-Wash contraints by one order of magnitude for $\lambda>{\rm a \, few\,}10^5$~m.}
\label{fig_yukawaB}
\end{figure}

\begin{figure}
\includegraphics[width=0.45\textwidth]{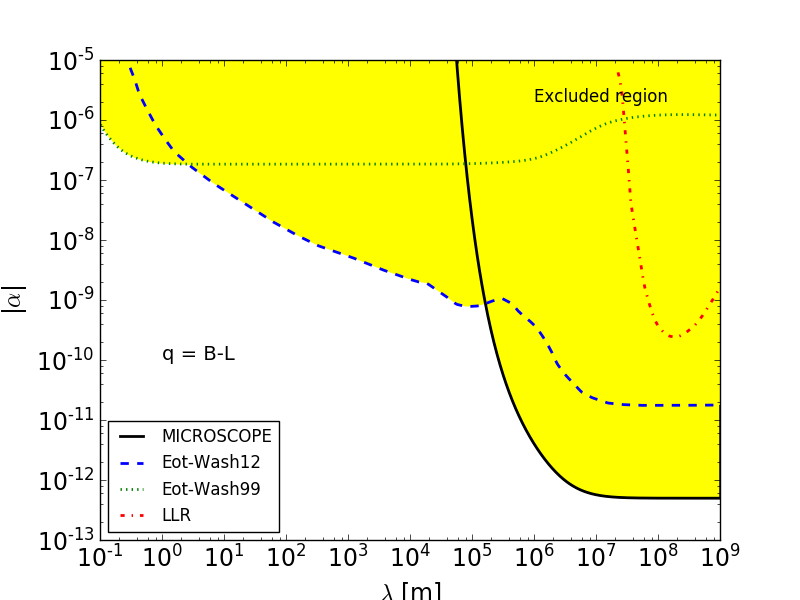}
\caption{Same as Fig. \ref{fig_yukawaB}, but with $q=B-L$, compared to the earlier constraints from Ref.~\cite{smith00} (dotted), Ref.~\cite{wagner12} (dashed) and Refs.~\cite{williams04, talmadge88} (dot-dashed).}
\label{fig_yukawaBL}
\end{figure}

\noindent{\it Baryonic/Leptonic charges}. The simplest analysis consists in assuming that the composition-dependent coupling $\alpha_{ij}$ depends on a scalar dimensionless ``Yukawa charge" $q$, characteristic of each material as~\cite{schlamminger08, wagner12}
\begin{equation}
\alpha_{ij} = \alpha \left(\frac{q}{\mu}\right)_i \left(\frac{q}{\mu}\right)_j,
\end{equation}
where $\alpha$ is a universal dimensionless coupling constant which quantifies the strength of the interaction with respect to gravity and $\mu$ is the atomic mass in atomic units (e.g. $\mu=12$ for carbon-12, or $\mu=47.948$ for titanium). Different definitions of the charge $q$ are possible depending on the detailed microscopic coupling of the scalar field to the standard model fields.  At the atomic levels, taking into account the electromagnetic and nuclear binding energies, the charge are usually reduced to the materials's baryon and/or lepton numbers ($B$ and $L$) (see e.g. Refs.~\cite{fayet90, fayet17}). Hence, for a macroscopic body, we must consider its isotopic composition. Hereafter, we shall set constraints on such interactions with either $q=B$ or $q=B-L$.

Following Ref.~\cite{touboul17a} and their approximations, it is straightforward to show (using Eqs. (\ref{eq_yukawa}) and (\ref{e.eta1})) that for MICROSCOPE, the E\"otv\"os parameter due to a Yukawa potential is

\begin{equation} \label{eq_eta}
\eta = \alpha \left[ \left(\frac{q}{\mu}\right)_{\rm Pt}\!\!\!\!\! - \left(\frac{q}{\mu}\right)_{\rm Ti}\right]\!\!\left(\frac{q}{\mu}\right)_E \left( 1 + \frac{r}{\lambda} \right) \hbox{e}^{-\frac{r}{\lambda}}
\end{equation}
where $r$ is the mean distance from the satellite to the center of the Earth \footnote{The orbit is nearly circular, and we can safely neglect any variation of $r$ (estimated less than 0.1\%).}. 
The Earth charge takes into account the Earth differentiation between core and mantle
\begin{equation} \label{eq_eta_qearth}
\left(\frac{q}{\mu}\right)_E = \left(\frac{q}{\mu}\right)_{\rm core} \Phi\left(\frac{R_c}{\lambda} \right) + \left(\frac{q}{\mu}\right)_{\rm mantle} \left[ \Phi\left(\frac{R_E}{\lambda} \right) - \Phi\left(\frac{R_c}{\lambda} \right)\right],
\end{equation}
where $R_E$ is the Earth mean radius and $R_c$ the Earth core radius.
The function  $\Phi(x) \equiv 3(x \cosh x - \sinh x)/x^3$ \cite{adelberger03} takes into account the fact that all Earth elements do not contribute similarly to the Yukawa interaction at the satellite's altitude \footnote{Note that we ignore the Earth's asphericity in this analysis.} ($\Phi=1$ for the test masses since their sizes are much smaller than the ranges $\lambda$ that can be probed in orbit). We assume that the core of the Earth is composed of iron and that the mantle is composed of silica (SiO$_2$) \cite{damour10a}. The baryonic and lepton charges for the MICROSCOPE experiment are summarized in Table \ref{table_q}.

At the 2-$\sigma$ level, MICROSCOPE's constraints on the E\"otv\"os parameter are given by Eq.~(\ref{e.etaMicro}), and can readily be transformed into constraints on Yukawa's ($\alpha$, $\lambda$). Figs~\ref{fig_yukawaB} and~\ref{fig_yukawaBL} depict the corresponding exclusion regions respectively for $q=B$ and $q=B-L$. In both analyses, we compare our new constraint to the bounds from E\"ot-Wash's torsion pendulum experiments~\cite{smith00, schlamminger08, wagner12} and the constraints from the Lunar-Laser Ranging (LLR) experiment~\cite{williams04, talmadge88}.  
Note that while we plot only the latest, most competitive constraints, several other experimental constraints are available (e.g. \cite{adelberger03, hoskins85, stubbs87, mitrofanov88, long99, hoyle01, hoyle04, kapner07, masuda09, sushkov11}).  Moreover, the LLR constraint could be slightly strengthened in the near future \cite{viswanathan17}.
This shows that MICROSCOPE's first results allow us to gain one order of magnitude compared to previous analyses for $\lambda > {\rm a \, few\, } 10^5$m. As MICROSCOPE orbits Earth at about 7000~km from its center, one would naively expect that it can only probe interactions with $\lambda > {\rm a \, few\, } 10^6 {\rm m}$; smaller ranges could not be probed as they imply too much of a damping at MICROSCOPE's altitude. However, if a fifth force with $\lambda \approx {\rm a \, few\,} 10^5 {\rm m}$ was strong enough to affect MICROSCOPE, the contribution from the nearest point of the Earth (as seen from MICROSCOPE) would be higher than that of the farthest point of the Earth, implying an asymmetric behavior that can be probed by MICROSCOPE (as captured by the function $\Phi(x)$ above). Hence, MICROSCOPE is sensitive to scalar interactions with ranges as low as a few hundreds of kilometers. \\

\noindent{\it Dilaton models}.  We now consider the characteristics of a generic dilaton with couplings described in Refs~\cite{damour94, damour10a, damour10b}.
The mass of an atom (atomic number $Z$ and mass number $A$) can be decomposed as $m(A,Z)=Zm_{\rm p}+(A-Z)m_{\rm n}+Zm_{\rm e} + E_1+E_3$ where $m_{\rm n,p}$ is the mass of the neutron or proton and $E_1$ and $E_3$ are the electromagnetic and strong interaction binding energies. Following Ref.~\cite{damour10a}, we consider that the coupling coefficients of the dilaton to the electromagnetic and gluonic fields are $d_e$ and $d_g$ while $d_{m_e}$, $d_{m_u}$ and $d_{m_d}$ are its coupling to the electron, $u$ and $d$ quarks mass terms. The latter two can be replaced by the couplings  $d_{\delta m}$ and $d_{\tilde m}$ to the symmetric and antisymmetric linear combination of $u$ and $d$. Assuming a linear coupling, one deduces that the variation of the fine structure constants and masses of the quarks are given by $\Delta\alpha_{\rm EM}/\alpha_{\rm EM}=d_e\phi/M_p$ and $\Delta m_{u,d}/m_{u,d}=d_{u,d}\phi/M_p$.

\begin{table}[t]
\caption{Baryonic, leptonic and dilaton charges for MICROSCOPE's test masses.}
\label{table_q}
\begin{ruledtabular}
\begin{tabular}{ccccc}
Material & $B/\mu$ & $(B-L)/\mu$ & $Q'_{\tilde{m}}$ & $Q'_e$ \\
\hline
Pt/Rh & 1.00026 & 0.59668 & 0.0859 & 0.0038 \\
Ti/Al/V & 1.00105 & 0.54044 & 0.0826 & 0.0019 \\
\end{tabular}
\end{ruledtabular}
\end{table}

\begin{figure}
\includegraphics[width=0.45\textwidth]{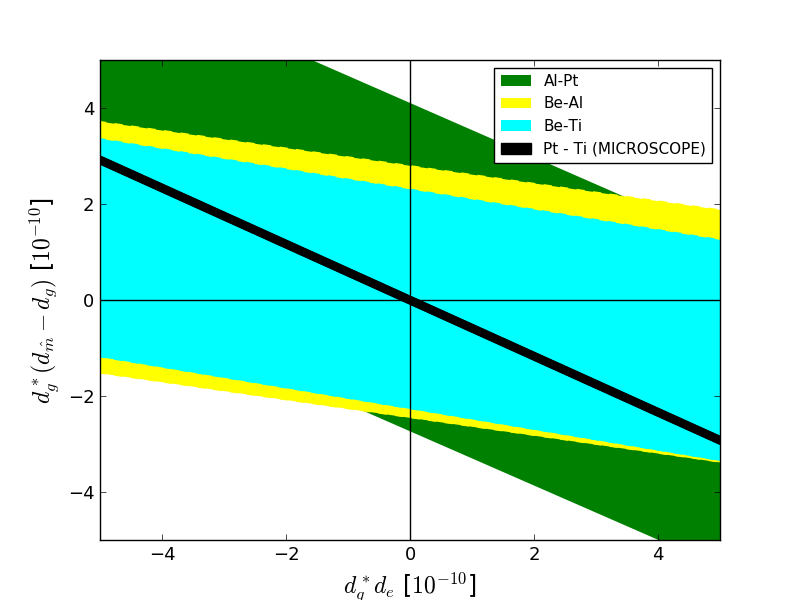}
\caption{Constraints on  the couplings of a massless dilaton $(D_{\tilde{m}},D_e$). The region allowed by the MICROSCOPE measurement (black band) is compared to earlier constraints by torsion pendulum experiments from Ref.~\cite{braginsky71} (green) and Ref.~\cite{wagner12} (yellow, cyan). The difference of slopes arises from the difference of material used in these 3 experiments. MICROSCOPE allows us to shrink the allowed region by one order of magnitude.}
\label{fig_masslessdilaton}       
\end{figure}

First, we consider a massless dilaton ($m_\phi =0$), whose range $\lambda_\phi$ is infinite, as was done by the E\"ot-Wash group \cite{wagner12}. The dilaton coupling to matter, and hence the fifth force, is parametrized by the 5 numbers ($d_g$, $d_e$, $d_{\tilde{m}}$, $d_{\delta m}$, $d_{m_e}$) so that the coupling to matter~(\ref{e.alphai}) takes the form
\begin{equation}
\alpha_i \approx d_g^* + \left[ \left( d_{\tilde m} - d_g \right) Q'_{\tilde{m}} + d_eQ'_e \right]_i,
\end{equation}
where $d_g^* = d_g + 0.093 (d_{\tilde{m}} - d_g) + 0.00027 d_e$. The dilaton charges depend on the chemical composition of the test masses and on the local value of the dilaton. Following Ref.~\cite{damour10a}, they are well-approximated by
\begin{equation}
Q'_{\tilde{m}} = 0.093-\frac{0.036}{A^{1/3}} - 1.4 \times 10^{-4} \frac{Z(Z-1)}{A^{4/3}}
\end{equation}
and
\begin{equation}
Q'_e = -1.4\times10^{-4}+7.7 \times 10^{-4} \frac{Z(Z-1)}{A^{4/3}}.
\end{equation}
In the limit where $\lambda$ is much larger than any other spatial scales, the E\"otv\"os parameter reduces to Eq.~(\ref{e.eta1}) so that (at first order in dilaton charges $Q'_j$ --given that $|Q'_j| \ll 1$) 
\begin{multline} \label{eq_eta_dilaton}
\eta_{\rm massless}   =  D_{\tilde{m}} \left([Q'_{\tilde{m}}]_{\rm Pt} - [Q'_{\tilde{m}}]_{\rm Ti} \right) + D_e \left( [Q'_e]_{\rm Pt} - [{Q'_e}]_{\rm Ti}\right),
\end{multline}
where the coefficients $D_{\tilde{m}} = d_g^* (d_{\tilde{m}} - d_g)$ and $D_e = d_g^* d_e$ are to be estimated. The values for $Q'_{\tilde{m}}$ and $Q'_e$ in the MICROSCOPE case are given in Table \ref{table_q}.

Fig.~\ref{fig_masslessdilaton} summarizes our new constraints and compare them to the earlier ones from the E\"ot-Wash \cite{wagner12} and the Moscow groups \cite{braginsky71}. The different slopes of the allowed regions are due to the different pairs of materials used by each experiment.
\\

\noindent{\it Massive dilaton}.  The mass of the dilaton modifies the range of its interaction so that Eq.~(\ref{eq_eta_dilaton}) is modified as
\begin{multline} \label{eq_eta_massive_dilaton}
\eta  =  \eta_{\rm massless} \times \Phi\left(\frac{R_E}{\lambda_\phi}\right) \left( 1+ \frac{r}{\lambda_\phi}\right) \hbox{e}^{-r/\lambda_{\phi}}.
\end{multline}

Note that this equation is simpler than Eq. (\ref{eq_eta_qearth}) because Eq. (\ref{eq_eta_dilaton}) does not depend on the Earth dilaton charge, and it is therefore independent of the exact Earth model used.

From Figs.~\ref{fig_yukawaB} and~\ref{fig_yukawaBL}, we expect that MICROSCOPE shall mainly be sensitive to masses in the range $10^{-14}-10^{-12}$~eV. Lower masses will result in constraints similar to those for a massless dilaton (see Fig. \ref{fig_masslessdilaton}) while larger masses cannot be constrained, as they correspond to ranges that MICROSCOPE cannot probe. This is indeed what we conclude from our analysis summarized in Fig.~\ref{fig_massivedilaton}. Constraints in the $(D_{\tilde{m}}, D_e$) plane are rather loose for high-enough masses,  $m_\phi > 10^{-12}$eV, and converge to those of a long-range dilaton for $m_\phi < 10^{-14}$eV.

\begin{figure}
\includegraphics[width=0.45\textwidth]{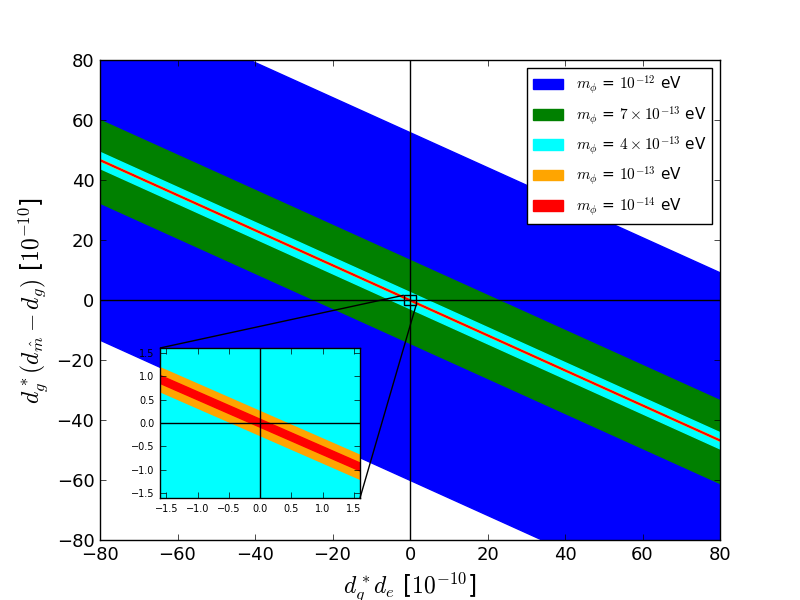}
\caption{Constraints on the couplings of a massive dilaton for various values of its mass. Each color shows the allowed $(D_{\tilde{m}}, D_e)$ for a given mass of the scalar field. The inset is a zoom on smaller $(D_{\tilde{m}}, D_e)$. Constraints saturate for light fields $m_\phi<10^{-14}$~eV. MICROSCOPE is not sensitive to masses larger than a few $10^{-12}$~eV.}
\label{fig_massivedilaton}       
\end{figure}

Finally, we assume that the dilaton  field couples only to the electromagnetic field, i.e. the only non-vanishing coupling is $d_e$. The coupling to proton and neutron is then induced from their binding energy~\cite{be1}. Several groups set constraints on such a dilaton from the fine structure constant oscillations in atomic frequency comparisons \cite{vantilburg15, hees16, kalaydzhyan17}. These results are based on the time evolution of the scalar field that oscillates within its self-potential. It has been argued that these oscillations may lead to oscillations of the Newtonian potentials if the scalar field behaves like cold dark matter \cite{blas17} (thereby affecting MICROSCOPE in an unexpected way), or even break the Yukawa approximation \cite{depirey17}. Here, we do not tie our scalar field to describe dark matter and we restrict our analysis to linear couplings, thence avoiding those possible pitfalls \footnote{see Supplemental Material}.
The MICROSCOPE constraints are obtained by considering the $D_{\tilde{m}}=0$-subspace of the parameter space $(D_{\tilde{m}}, D_e, m_\phi)$ of Fig.~\ref{fig_massivedilaton}, and recognizing that $D_e=d_g^*d_e=0.00027d_e^2$. Fig.~\ref{fig_de} shows our constraints, compared with those from the E\"ot-Wash test of the WEP and with atomic spectroscopy \cite{vantilburg15, hees16}. MICROSCOPE allows us to exclude a new region above $|d_e|=10^{-4}$, for a field of mass $10^{-18} < m_\phi/{\rm eV}<10^{-11}$. Atomic spectroscopy stays more competitive for lighter fields. \\

\begin{figure}
\includegraphics[width=0.45\textwidth]{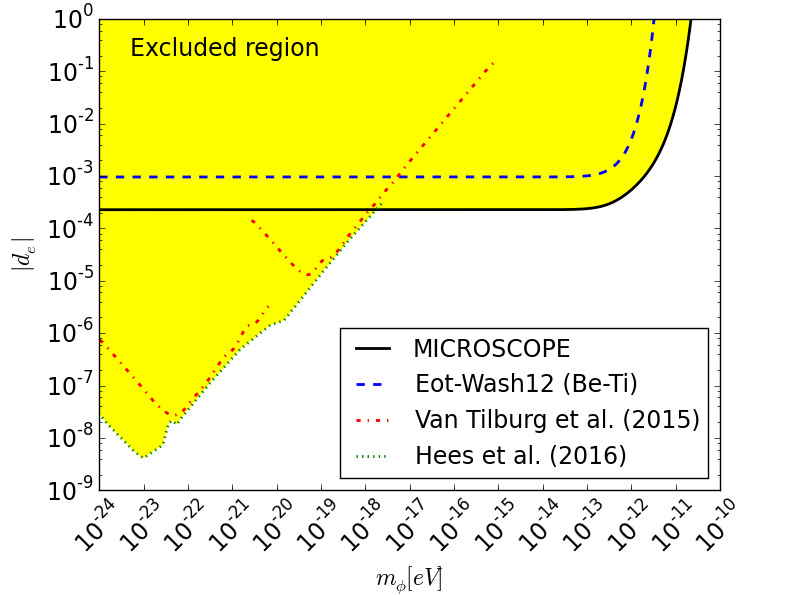}
\caption{Constraints on $d_e$, for a dilaton coupled only to the electromagnetic sector, compared with constraints from atomic spectroscopy (dot-dashed \cite{vantilburg15, hees16}) and E\"ot-Wash WEP test (dashed \cite{schlamminger08}).}
\label{fig_de}       
\end{figure}

\noindent{\it Conclusion}.  This letter gave the first constraints on a composition-dependent scalar fifth force from MICROSCOPE's first measurement of the WEP \cite{touboul17a}. 
We first considered the case of a massive scalar field coupled to either $B$ or $B-L$ to conclude that MICROSCOPE is particularly competitive for a Yukawa potential of range larger than $10^5$m (corresponding to a field of mass smaller than $10^{-12}$eV). In that case, we  improved existing constraints on the strength of the field by one order of magnitude. Below that range, torsion pendulum experiments remain unbeaten. Then, we considered a model describing the coupling of a generic dilaton to the standard matter field with 5 parameters, both for a massless and massive field. For $m_\phi < 10^{-14}$~eV, our constraints are similar to those for a massless field and better by one order of magnitude than the previously published ones.

From a theoretical perspective, a scalar long-range interaction is severely constrained by its effects on planetary motion. Since general relativity passes all tests on Solar-System scales many mechanisms have been designed to hide this scalar field in dense regions (e.g. chameleons \cite{khoury04a, khoury04b}, symmetron \cite{hinterbichler10}, K-mouflage \cite{babichev09, brax13} or Vainshtein \cite{vainshtein72}). The generic dilaton model considered in this letter corresponds to another type of screening (the least coupling principle \cite{damour94}) and can incorporate the behavior of many theories, such as string theory. The local prediction of the violation of the WEP can be compared to the variation of the fundamental constants on local and astrophysical scales (e.g. \cite{Luo, guena12, leefer13, tobar13}). Better constraints can be obtained from modeling  the profile (and time variation) of the scalar field along MICROSCOPE's orbit, as well as its propagation inside the satellite up to the test masses; this is non-trivial, requires some care, and will be done in a further work. Constraints on the violation of the WEP  will also  have strong consequences for bigravity models~\cite{bius}.

From an experimental perspective, these new constraints were obtained from only two MICROSCOPE's measurement sessions of the E\"otv\"os parameter \cite{touboul17a}. As the mission is scheduled to continue until 2018, new data are currently coming in, thereby offering the possibility of decreasing the statistical errors. We are also refining our data analysis procedures to optimize the measurement of the WEP. We therefore expect to improve on MICROSCOPE's constraint on the E\"otv\"os parameter by the end of the mission: ten times as many data will be available than were used in Ref.\cite{touboul17a}; furthermore, although we expect the data to become systematic-dominated, the control on systematics will be improved compared to Ref. \cite{touboul17a}, since calibration sessions have been performed, whose results will be used in the next data analysis. 
Therefore, we could improve the constraints reported in that letter by up to another order of magnitude (unless a WEP violation becomes apparent). But this forecast is valid only for $\lambda>{\rm a \, few \,}10^5$~m ($m_\phi<10^{-12}$~eV). Probing lower-range (more massive) scalar fields can be done only using small scale experiments. Torsion pendulum and atomic interferometry experiments represent our best hopes to look for such extra-fields. New, improved torsion pendulum will then be required to probe laboratory and smaller scale gravity, either through the measurement of the WEP or of the gravitational inverse square law. A torsion pendulum experiment in space seems the way forward to beat the current on-ground limits \cite{moriond}. \\

\noindent{\it Acknowledgements}.  We thank the members of the MICROSCOPE Science Performance Group for useful discussions. We also thank Thibault Damour for useful comments on a first version of the manuscript. This work makes use of technical data from the CNES-ESA-ONERA-CNRS-OCA Microscope mission, and has received financial support from ONERA and CNES. We acknowledge the financial support of the UnivEarthS Labex program at Sorbonne Paris Cit\'e (ANR-10-LABX-0023 and ANR-11-IDEX-0005-02). The work of JPU is made in the ILP LABEX (under reference ANR-10-LABX-63) was supported by French state funds managed by the ANR within the Investissements d'Avenir programme under reference ANR-11-IDEX-0004-02. This work is supported in part by the EU Horizon 2020 research and innovation programme under the Marie-Sklodowska grant No. 690575. This article is based upon work related to the COST Action CA15117 (CANTATA) supported by COST (European Cooperation in Science and Technology).

\bibliography{micyd}

\clearpage
\newpage
\maketitle
\onecolumngrid
\begin{center}
\textbf{\large MICROSCOPE mission: first constraints on the violation of the weak equivalence principle by a light scalar dilaton} \\ 
\vspace{0.05in}
{ \it \large Supplemental Material}\\ 
\vspace{0.05in}
{Joel Berg\'e, Philippe Brax, Gilles M\'etris, Martin Pernot-Borra\`as, Pierre Touboul, Jean-Philippe Uzan}
\end{center}
\setcounter{equation}{0}
\setcounter{figure}{0}
\setcounter{table}{0}
\setcounter{section}{0}
\makeatletter
\renewcommand{\theequation}{S\arabic{equation}}
\renewcommand{\thefigure}{S\arabic{figure}}
\renewcommand{\thetable}{S\arabic{table}}
\newcommand\ptwiddle[1]{\mathord{\mathop{#1}\limits^{\scriptscriptstyle(\sim)}}}




We are interested in scalar-tensor theories of the type
\begin{equation}
S= \int d^4x \sqrt{-g} \left ( \frac{R}{16\pi G_N} -\frac{1}{2} (\partial \phi)^2 -\frac{m^2}{2} \phi^2\right ) 
+\sum_i S_m( A^2_i(\phi) g_{\mu\nu})
\end{equation}
where different particle species $i$ couple to matter via the coupling function $i$. In the core of the paper we also consider the coupling to gauge fields in the form of
\be
S_F= -\int d^4x \sqrt{-g} \frac{B_i(\phi)}{4} F^{\mu\nu}_i F_{\mu\nu}^i
\ee
which plays a role to give the gluon and electromagnetic parts of the atomic masses. In this section we focus on the role of the coupling $A_i$ whilst the generalisation which includes
the $B_i$'s is taken into account in the text following Damour and Donoghue \cite{damour10b}, and Damour and Polyakov \cite{damour94}.

The scalar field is a canonically normalised scalar field with a mass $m$. We focus on linearly coupled scalars where
\be
A_i(\phi)= e^{\alpha_i \phi/\sqrt 2 m_{\rm Pl}}
\label{linear}
\ee
where here and contrarily to the main text $m_{\rm pl}^2 = 1/8\pi G_N$.
This choice is crucial as if the potential were non-linear such as the inverse power law $V(\phi) \sim 1/\phi^n$ for instance, the effects of the scalar  field on matter could be screened, see the review \cite{brax17}
and references therein. If the coupling function were universal $A_i\equiv A$ and of the scalarisation type $A(\phi)=e^{\beta \phi^2/2 m_{\rm Pl}^2}$ as in \cite{depirey17}, the solutions
to the Klein-Gordon equation in the presence of matter would be destabilised due to the negative coupling $\beta <0$ and lead to a completely different phenomenology.
Here we restrict ourselves to the linear couplings (\ref{linear}) leading to a Yukawa interaction between massive objects as we recall below.

Notice too that we do not tie our scalar field to describe dark matter. Indeed, a massive scalar oscillating around the minimum of its potential behaves like Cold Dark Matter
over time scales larger than the oscillation time. This is what is assumed in \cite{blas17} where the oscillations of the coupled scalar lead to oscillations of the Newtonian potentials, which could become
observable in the change of periods of binary pulsars. In our case, the field can be assumed to store an energy in the form of oscillations which is negligible compared to
the dark matter energy density. This can be easily realised if for instance the scalar field is present during inflation and its amplitude decays exponentially fast in $a^{-3/2}$ towards zero, in such a way
that the field is static at the end of inflation.

Let us recall how Eq. (2) of the main text is derived, see \cite{hui09} that we adapt to the present situation.
First of all let us recall that  the Einstein equation reads
\be
G_{\mu\nu}= 8\pi G_N (T^m_{\mu\nu}+ T^\phi_{\mu\nu})
\ee
where the energy momentum of matter is given by
\be
T^m_{\mu\nu}= A(\phi) \rho u_\mu u_\nu
\ee
where $\rho$ is the conserved matter density and $u_\mu$ the velocity 4-vector of matter.
Working in the Newtonian gauge for the metric
\be
ds^2=-(1+2\Phi_N) dt^2 + (1-2\Phi_N)dx^2
\ee
 we can expand both the scalar field and the Newtonian potential as
\be
\Phi_N= \bar \Phi +\delta \Phi,\ \phi= \bar\phi +\delta \phi
\ee
where a bar denotes a background quantity and  ``$\delta$'' the fields sourced by  objects. In the vicinity of an object, the background field can be expanded to linear order
\be
\bar\Phi(\vec x)= \bar \Phi_0 +\partial_i \bar \Phi (\vec x) x^i, \ \bar\phi(\vec x)= \bar\phi_0 +\partial_i \bar\phi (\vec x) x^i.
\ee
The field outside a given object $B$ and created by this object satisfies the static Klein-Gordon equation, where here we assume that the size of the objects is much smaller than the range
$1/m$ of the scalar interaction, 
\be
\Delta \delta \phi - m^2 \delta \phi=\frac{\alpha_B m_B }{\sqrt 2 m_{\rm Pl}} \delta^{(3)}(\vec x)
\ee
and the matter density has been approximated by a point-like source. 
The  solution is the Yukawa potential
\be
\delta  \phi= - \frac{\alpha_B m_B }{4\pi\sqrt{2} m_{\rm Pl} r}e^{-m r}.
\ee
where $r=\sqrt{\vec x^2}$.
For extended objects, we need to integrate this point-like solution, which leads to the shape function of the main text's Eq. (7).
We also assume that the object creates the Newtonian potential
\be
\delta\Phi(r) = -\frac{G_N m_B}{r}.
\ee
 We assume that matter is responsible for the Newtonian potential, and that the scalar field energy scale is negligible compared to matter inside the object
and very small outside. This is the case for instance if the scalar field sits at the minimum of the potential $\bar \phi_0=0$.

The Einstein equation can be rewritten as
\be
G^{(1)}_{\mu\nu}= 8\pi G_N (T^m_{\mu\nu}+ T^\phi_{\mu\nu}+  t_{\mu\nu}).
\ee
where $G^{(1)}_{\mu\nu}$ is linear in the Newtonian potential and the pseudo-tensor is given by
\be
t_{\mu\nu}=-\frac{1}{8\pi G_N} G^{(2)}_{\mu\nu}
\ee
where $G^{(2)}_{\mu\nu}$ contains all the higher order terms in the Newtonian potential.
The mass $m_B$ of the object becomes
\be
m_B= -\int_V d^3x \tilde T_0^0.
\ee
where we  draw a sphere of volume $V$ encircling the object and
\be
\tilde T_{\mu\nu}=T^m_{\mu\nu}+ T^\phi_{\mu\nu} + t_{\mu\nu}.
\ee
Neglecting the scalar contribution to the energy density, the mass is given by the integral over the object
\be
m_B= \int_V A(\phi) \rho d^3x.
\ee
The momentum of the object is then
\be
P^B_i= \int_V  d^3 x \tilde T_i^0.
\ee
Using the  non-covariant Bianchi identity which  implies that $\partial^\mu \tilde T_{\mu\nu}=0$, we find that
\be
\dot P^B_i= - \int_{\partial V} dS_j \tilde T^j_i
\ee
where the surface integral is on the surface of the outer sphere. There the matter energy momentum tensor is negligible, and similarly for the contribution from the scalar field energy density. Only two terms have a relevant flux: the scalar and gravitational ones.
The gravitational flux has been computed in \cite{hui09} and yields
\be
\int_{\partial V} dS_j t^j_i= m_B \partial_i \bar \Phi.
\ee
The new contribution from the scalar field is simply dominated by the large gradient of the scalar field $\delta \phi(r)$
\be
-\int _{\partial V} dS_j T^{\phi j}_i
=- \frac{\alpha_B m_B}{\sqrt 2 m_{\rm Pl} } \partial_i \bar\phi \,.
\ee
As a result we obtain that
\be
\dot P^B_i= -m_B \partial_i \bar\Phi-\frac{\alpha_B m_B}{\sqrt 2 m_{\rm Pl}}\partial_i \bar\phi.
\ee
Similarly the centre of mass coordinates satisfies the modified Newton equation
\be
\ddot X_B^i= -\partial^i \bar\Phi - \alpha_B \frac{\partial^i \bar\phi}{\sqrt 2 m_{\rm Pl}}.
\ee
When the external fields $\bar\Phi$ and $\bar\phi$ are due to another extended object, we have
\be
\bar\Phi= -\frac{G_N m_A}{\vert \vec x -\vec x_A\vert}, \ \ \bar\phi= - \frac{\alpha_A m_A }{4\pi\sqrt{2} m_{\rm Pl}\vert \vec x -\vec x_A\vert }e^{-m \vert \vec x -\vec x_A\vert}.
\ee
The motion of the object $B$ is due to the total potential
\be
\Phi_{AB}=
(1+\alpha_A \alpha_B e^{-m\vert \vec x -\vec x_A\vert})\Phi_A
\ee
where $\Phi_A$ is the Newtonian potential due to the second object $A$ and  we have
\be
\ddot X^i_A= - \partial^i \Phi_{AB}.
\ee

\end{document}